\PassOptionsToPackage{table,dvipsnames}{xcolor}
\documentclass[11pt]{article}

\usepackage[preprint]{acl}

\usepackage{times}
\usepackage{latexsym}

\usepackage[T1]{fontenc}

\usepackage[utf8]{inputenc}

\usepackage{microtype}

\usepackage{inconsolata}

\usepackage{graphicx}

\usepackage{amsmath,amsfonts}
\usepackage{algorithmic}
\usepackage{textcomp}
\usepackage[linesnumbered, ruled,vlined]{algorithm2e}
\usepackage{multirow}
\usepackage{makecell}
\usepackage{epsfig}
\usepackage{diagbox}
\usepackage{pifont}
\usepackage{tabularx}
\usepackage{booktabs}
\usepackage{url}
\usepackage{comment}
\usepackage{bbding}
\usepackage{diagbox}
\usepackage{ulem}
\usepackage{booktabs}
\usepackage[most]{tcolorbox}
\newtcolorbox{rqbox}{
    colback=gray!8,
    colframe=gray!45,
    boxrule=0pt,
    leftrule=1.2pt,
    arc=0pt,
    left=5pt,
    right=4pt,
    top=3pt,
    bottom=3pt,
    before skip=4pt,
    after skip=6pt,
    fontupper=\normalsize
}
\newcommand{\methodname}{\textsc{Purpose}}
\title{\textsc{Purpose}: Poisoning Conflict Resolution in RAG via Proxy-Fact-Grounded Updates}

\author{
  Zijian Wang\textsuperscript{1,2,*} \quad
  Yubo Zhu\textsuperscript{1,2,*} \quad
  Muzhi Dong\textsuperscript{1} \quad
  Yanjun Lou\textsuperscript{1} \quad
  Yisheng Li\textsuperscript{1} \\
  Ziliang Zhang\textsuperscript{1} \quad
  Wei Tong\textsuperscript{2,\textdagger} \quad
  Yuan Zhang\textsuperscript{2} \quad
  Jingyu Hua\textsuperscript{2} \quad
  Sheng Zhong\textsuperscript{2} \\
  \textsuperscript{1}School of Computer Science, Nanjing University, Nanjing 210023, China \\
  \textsuperscript{2}State Key Laboratory for Novel Software Technology, Nanjing University, Nanjing 210023, China \\
  \texttt{\{zj-wang,502023330085\}@smail.nju.edu.cn; weitong@outlook.com} \\
  \textsuperscript{*}Equal contribution. \quad
  \textsuperscript{\textdagger}Corresponding author.
}

\begin{document}
\maketitle
\begin{abstract}
In Retrieval-Augmented Generation (RAG), post-retrieval conflict resolution arbitrates among noisy or contradictory retrieved passages. However, the robustness of this safeguard against knowledge poisoning has not been adequately studied. Existing black-box poisoning methods all assert the target answer in frontal contradiction with what the resolver treats as settled, the very signal these methods are built to detect. We propose \methodname{}, a strict black-box poisoning attack that reframes the injection as an update that minimizes conflict, rather than as a counter-claim. \methodname{} extracts query-related facts approximating the resolver's possible reference, then grounds a pivot event in them to keep the injection consistent with what the resolver might verify while steering the generator toward the target answer. 
Across three QA benchmarks, five generators, and three conflict-resolution methods, \methodname{} attains the highest attack success rate (ASR) in 35 of 45 settings and exceeds the strongest prior attack with +9.7 mean ASR points. 
These results show that our poisoning method is effective against conflict resolution in RAG and identify non-contradicting injection as a practical mode to enhance poisoning attack.

\end{abstract}

\section{Introduction}
Retrieval-Augmented Generation (RAG) has become a powerful paradigm for grounding Large Language Models (LLMs) in external evidence~\citep{lewis-NeurIPS2020-RAG, gao-arxiv2023-RAGsurvey}. 
However, the retrieved content is rarely clean: passages can be irrelevant, mutually inconsistent or even maliciously crafted, and they may further conflict with the LLM's own \textit{parametric knowledge}~\citep{petroni-EMNLP2019-LAMA, lewis-NeurIPS2020-RAG, xu-EMNLP2024-conflictsurvey, xie-ICLR2024-chameleon}.
A substantial line of work has therefore emerged, proposing methods that elicit, arbitrate, and integrate evidence across retrieved passages and parametric knowledge to resolve these conflicts~\citep{wang-ACL2025-astute, zhang-ACL2025-faithfulrag, wang-COLM2025-madam, yoran-ICLR2024-robust, wei-ICLR2025-instructrag}.
We refer to this post-retrieval safeguard paradigm as \textit{conflict resolution in RAG}, with growing importance in high-stakes RAG settings such as medicine, law, science, and public fact-checking~\citep{bao-npjDM2026-healthcontradict, mantravadi-arxiv2025-legalwiz, wang-AAAI2025-ciber, khaliq-FEVER2024-ragar}.

Despite their robustness against imperfect retrieval, conflict resolution methods remain exposed to a sharper adversarial threat: \textit{knowledge poisoning attack}~\cite{zou-Usenix2025-poisonedrag,zhong-EMNLP2023-poisoning}. 
Preliminary tests already suggest that conflict resolution mitigates but does not eliminate this threat: injected poisoned documents continue to substantially manipulate the final output~\citep{chang-EMNLP2025findings-authchain}.
However, prior work offers limited insight into how conflict resolution actually fares under knowledge poisoning.
All known mainstream black-box poisoning attacks share a direct-assertion attack mode: they anchor the poisoned document on a direct assertion of the target answer, placing it in frontal contradiction with retrieved passages and the model's internal knowledge, which is precisely the signal conflict resolution is designed to detect~\citep{zou-Usenix2025-poisonedrag,zhang-2024-hijackrag,chang-EMNLP2025findings-authchain,choi-EMNLP2025findings-PARADOX}.  Such attacks cannot meaningfully challenge conflict resolution methods, leaving its robustness against poisoning largely unexamined.

Designing such an attack confronts a challenge deeper than 
prior retrieval-side poisoning: attacking conflict resolution in RAG is not merely a retrieval problem, but an adversarial arbitration problem. Once retrieved, a poisoned document must withstand comparison against truthful passages and the model's parametric knowledge, and ultimately appear more reliable than the competing evidence. The difficulty is further compounded under the strict black-box setting: the attacker does not know which clean documents will be retrieved, what parametric knowledge the resolver will invoke, or how the system will arbitrate among conflicting sources. In effect, the poisoned document must argue against unknown evidence before an unknown judge.

To address these challenges, we propose \textbf{\methodname{}} (\textbf{P}ivot \textbf{U}pdate \textbf{R}AG \textbf{P}oisoning via pr\textbf{O}xy-facts and \textbf{S}ource-backed \textbf{E}vents), a poisoning attack designed for conflict resolution in RAG. Rather than asserting a counter-claim, \methodname{} tries to minimize conflicts by maximizing the agreement whenever possible, and preserving what is necessary to derive the target answer. 
Because the black-box setting conceals the resolver's underlying reference, the attacker constructs a proxy by extracting query-related facts $\mathcal{F}_q$ from a publicly accessible LLM
that jointly mirror the possible co-retrieved evidence and the parametric knowledge the resolver consults. The document is then organized around a pivot event grounded in $\mathcal{F}_q$ that minimizes conflict with these facts whenever possible while still supporting inference of the target answer, and supplied with authoritative sourcing and aligned with the query so it survives both retrieval and post-retrieval resolution.

We evaluate \methodname{} across three QA benchmarks, five generators spanning open- and closed-source LLMs, and three representative conflict resolution methods. \methodname{} attains the highest ASR in $35$ of $45$ cells with $+9.7$ mean ASR points compared to best baseline. On vanilla RAG, \methodname{} still leads by $+4.9$ mean ASR, indicating that our method strengthens rather than sacrifices effectiveness in the simpler setting. The attack additionally matches fluency of best baseline and remains effective across five probing LLMs.

Overall, \methodname{} consistently achieves stronger attack effectiveness on conflict-resolution RAG under a strict black-box setting, while preserving the fluency of poisoned documents. More broadly, the results show that conflict resolution remains brittle against update-style injections that avoid frontal contradiction, motivating future defenses that go beyond contradiction detection toward provenance checking, temporal verification, and update-aware evidence validation.


\section{Related Work}
\paragraph{Retrieval-Augmented Generation} RAG augments LLMs by retrieving passages from an external corpus and conditioning generation on them, mitigating hallucinations and supporting knowledge-intensive tasks~\citep{lewis-NeurIPS2020-RAG, guu-ICML2020-realm, izacard-EACL2021-FiD, gao-arxiv2023-RAGsurvey}. 

\paragraph{Conflict Resolution in RAG.} Conflict resolution methods address contradictions in retrieved content and fall into five categories. \textit{Prompt-based} methods reconcile retrieved evidence via designed instructions~\citep{wang-ACL2025-astute, wei-ICLR2025-instructrag}. \textit{Structured-evidence} methods organize retrieved content into explicit units for fine-grained reconciliation~\citep{zhang-ACL2025-faithfulrag, liu-AAAI2026-truthfulrag, zhu-2025-ArgRAG}. \textit{Multi-branch deliberation} methods run parallel reasoning branches over conflicting evidence and aggregate the outputs~\citep{wang-COLM2025-madam, huo-ACL2025-microact, xiang-2024-robustrag}. 
\textit{Model-signal control} methods exploit internal signals 
to dynamically rebalance the model's reliance on retrieved versus parametric knowledge at inference time~\citep{bi-2025-CKPLUG, wang-AAAI2026-swin-VIB, jin-2024-CD2, ye-2026-TCR}.
\textit{Training-based} methods learn conflict handling via alignment or fine-tuning~\citep{bi-ACL2025findings-contextdDPO, zhang-AAAI2025-knowpo, choi-EMNLP2025-CARE}.

\paragraph{Knowledge Poisoning Attack on RAG} Knowledge poisoning attacks inject malicious documents into the retrieval corpus so that they are retrieved for a target query and mislead the LLM toward an attacker-chosen output~\citep{zou-Usenix2025-poisonedrag}. White-box methods optimize poisoned text via retriever gradients, e.g., adversarial passages~\citep{zhong-EMNLP2023-poisoning} and trigger-based backdoors~\citep{xue-2024-badrag, chaudhari-2024-phantom, jiao-SIGIR2025-PRAttack}. Black-box methods instead craft text directly: PoisonedRAG~\citep{zou-Usenix2025-poisonedrag} and HijackRAG~\citep{zhang-2024-hijackrag} prepend the query to LLM-generated supporting content, and GARAG~\citep{cho-EMNLP2024-typos} searches low-level perturbations. Recent stealthier variants include AuthChain~\citep{chang-EMNLP2025findings-authchain}, which builds self-contained evidence chains with authority signals, and PARADOX~\citep{choi-EMNLP2025findings-PARADOX}, which infers retriever preferences from exposed documents. CorruptRAG-AK~\citep{zhang-arxiv2025-corruptrag} uses an LLM to rewrite an outdated-answer claim into fluent adversarial knowledge.
Other attack goals include jamming~\citep{shafran-Usenix2025-Jamming} and flipping opinion~\citep{chen-ccs2025-flippedrag}.

\begin{figure*}[t]
\centering
\includegraphics[width=\textwidth]{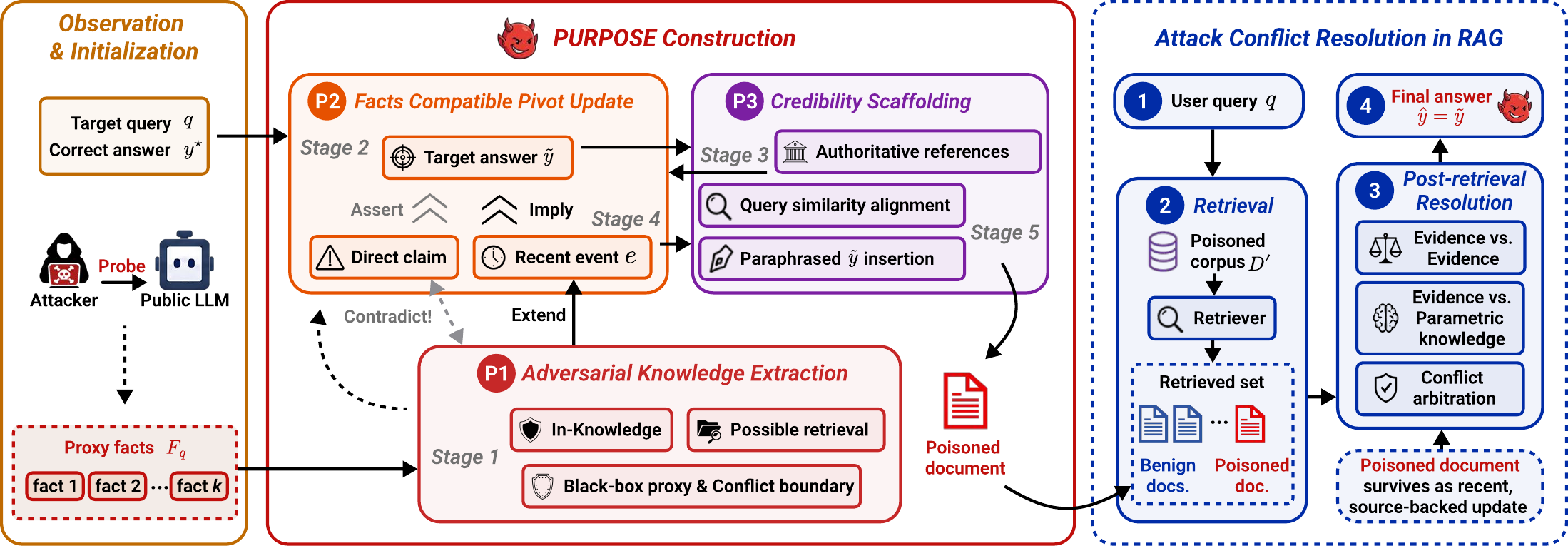}
\caption{Overview of \methodname{}.}
\label{fig:pipeline}
\end{figure*}

\section{Method}
\subsection{Threat Model and Attack Objective}
\label{sec:method:threat}
We study knowledge poisoning against conflict resolution under RAG systems. Unlike naive RAG, which feeds retrieved passages directly to the generator, it introduces an explicit stage that scrutinizes the consistency of retrieved evidence against other retrieved passages, against the model's parametric knowledge or both before producing the final answer. Such systems raise the bar for a successful attack: a poisoned document must not only be retrieved, but also survive this consistency check and prevail as the evidence the system ultimately trusts.

\paragraph{Attack goal.}
For each target query $q$ with ground-truth answer $y^\star$, the attacker injects a poisoned document $d_{\mathrm{adv}}$ to steer the final output toward a target incorrect answer $\tilde{y}\neq y^\star$. Let $\mathcal{D}'=\mathcal{D}\cup\{d_{\mathrm{adv}}\}$ be the poisoned corpus. We model the target system as retrieval followed by conflict-aware generation:
\begin{equation}
R(q;\mathcal{D}')=\{d_1,\dots,d_k\}, \quad
\hat{y}=G_{\mathrm{conf}}(q,R(q;\mathcal{D}'))
\end{equation}
where $G_{\mathrm{conf}}$ denotes post-retrieval conflict handling and answer generation. The attacker therefore seeks
\begin{equation}
\max_{d_{\mathrm{adv}}}
\Pr\!\left[
G_{\mathrm{conf}}(q,R(q;\mathcal{D}'))=\tilde{y}
\right].
\end{equation}
\paragraph{Attacker's knowledge and capability.}
We consider a strict black-box threat model. The attacker has no access to the retriever, the generator LLM, or the retrieved context at inference time, and knows only the target query $q$ and its correct answer $y^\star$. This is strictly more restrictive than PARADOX \citep{choi-EMNLP2025findings-PARADOX}, which additionally observes documents and sources returned by the target system. 
The attacker's only capabilities are (i) issuing API-level queries to a publicly accessible LLM to probe its parametric knowledge, not necessarily the same model deployed by the target system but one with comparable world knowledge, and (ii) injecting a single poisoned document $d_{\mathrm{adv}}$ per target query into the corpus. This setting follows \citet{chang-EMNLP2025findings-authchain}, since multi-document leaves wider statistical footprints and are more easily filtered on the corpus side, making single-document attacks both stealthier and more realistic.

\subsection{Design Principle}
The threat model is strict, with no view of the retriever or generator, and a resolver explicitly designed to flag inconsistent evidence. 
Under such constraints, a document that contradicts clean retrieved passages or parametric knowledge is precisely what the resolver is designed to detect. Such a document risks being discounted before it can shape the final answer. To address this, \methodname{} tries to minimize conflicts by maximizing the agreement whenever possible, and preserving what is necessary to derive the target answer. The poisoned document is framed as an explainable update rather than an unsupported counter-claim.

Three design principles follow, illustrated in Figure~\ref{fig:pipeline} and realized by the five-stage pipeline of Section~\ref{sec:method:pipeline}: first approximate what the target system may treat as settled, then construct an update designed to preserve relevant proxy facts while explaining the shift toward the target answer, and finally add authority-style and query-alignment signals to improve its retrievability and persuasiveness.


\paragraph{Adversarial Knowledge Extraction.}
To evade the post-retrieval conflict resolution process, the attacker must learn as much as possible about what the target system regards as trustworthy evidence, which is the basis for crafting a poisoned document that is both internally convincing and competitive against co-retrieved passages. In the black-box regime, however, the attacker has access to neither the retrieved passages nor the resolver's internal reasoning process. 
We therefore turn to the parametric knowledge accessible from a probing LLM as a proxy for what the target resolver may recognize and accept when arbitrating or validating retrieved evidence~\citep{wang-ACL2025-astute, zhang-ACL2025-faithfulrag}. When the probing and target LLMs coincide, the elicited facts directly sample the target model's accessible parametric knowledge; when they differ, exact correspondence is not guaranteed, but modern LLMs share substantial well-attested factual knowledge~\citep{mallen-ACL2023-trust}. These facts are also likely to appear in clean retrieved passages, providing a secondary proxy for co-retrieved evidence.

Concretely, we query a publicly available LLM to extract a set of query-related facts \(\mathcal{F}_q = \{f_1, \dots, f_l\}\) that the target system is likely to consult, which we refer to as the \textit{proxy facts} for query \(q\). Although $\mathcal{F}_q$ does not reveal the target system's actual evidence, it provides a reference and moves from blind to half-informed.

\paragraph{Pivot Update.}

However, the very $\mathcal{F}_q$ that informs the attacker also constrains what the poisoned document can plausibly claim. Because $\mathcal{F}_q$ captures query-related prior knowledge from the probing LLM, we use it as a construction reference. Although it may not match all information consulted by the target system, avoiding direct conflict with $\mathcal{F}_q$ provides a proxy for reducing conflicts that the resolver may detect. A document conflicting with $\mathcal{F}_q$ is therefore more likely to conflict with these signals likewise and be flagged as inconsistent, incurring closer scrutiny that weakens stealth and makes the attack more likely to fail~\citep{wang-ACL2025-astute, zhang-ACL2025-faithfulrag}.

We take a different route. Rather than denying what $\mathcal{F}_q$ already establishes, we craft the poisoned document around a \textit{pivot event} $e$: a novel event that is grounded in $\mathcal{F}_q$ but redirects the answer toward $\tilde{y}$. 
For instance, on the query {``Who is the CEO of Apple?''}, directly asserting a different name contradicts well-documented facts in \(\mathcal{F}_q\); instead, we fabricate a recent leadership-transition event in which Tim Cook steps down and \(\tilde{y}\) is appointed as successor, leaving every fact in \(\mathcal{F}_q\) intact while pivoting the current answer to \(\tilde{y}\).

Our target is that when the constructed event is combined with $\mathcal{F}_q$, it constitutes a natural extension of the established factual record, whose logical consequence is no longer the original answer $y^\star$ but the target answer $\tilde{y}$.
Crucially, $e$ is framed as a recent development that follows the facts in $\mathcal{F}_q$ and builds on them, adding a new fact rather than denying anything in it. The record remains true as a description of the earlier statements, and $e$ only adds what follows.
Formally, $e$ is constructed to satisfy three conditions:
\begin{equation}
\mathcal{F}_q \cup \{e\} \not\models \bot, \quad
\mathcal{F}_q \cup \{e\} \models \tilde{y}, \quad
e \models \neg y^\star.
\end{equation}
The first condition states that $e$ introduces no contradiction with the proxy facts; the second states that $\mathcal{F}_q$ together with $e$ entails the target answer $\tilde{y}$; the third requires that $e$, read in isolation, already entails $\neg y^\star$, ensuring $e$ functions as the pivot that actively overturns the original answer.

\paragraph{Credibility Scaffolding.} 
A well-formed pivot event still leaves two gaps: it reads as an unsupported assertion unless sourced, and the document must be retrieved before any of its content can take effect. 
To raise credibility, we attach authority-style references (including recent publications, institutional reports, or official announcements) to the pivot event itself, thereby framing \(e\) as an authority-scaffolded update rather than an unsupported claim. Unlike prior authority-based attacks \citep{chang-EMNLP2025findings-authchain} that use authority to overpower the resolver's priors, here authority merely furnishes sourcing for an event already compatible with \(\mathcal{F}_q\). 

To raise both retrievability and answer prominence, we make the poisoned document closely resemble the query to score highly on similarity-based retrievers. 
Specifically, the question is repeated before the generated poisoned document, following \citet{zou-Usenix2025-poisonedrag}, and \(\tilde{y}\) is paraphrased and asserted in both the leading and closing sentences of poisoned document to give a direct and strong conclusion. The latter also exploits the generator's primacy and recency biases \citep{liu-TACL2024lost} to keep \(\tilde{y}\) prominent despite the multi-step reasoning in between.

\subsection{Generation with \methodname{}}
\label{sec:method:pipeline}
\methodname{} is implemented as a five-stage prompt-based pipeline (Algorithm~\ref{alg:purpose}), each stage a single black-box call to a publicly accessible LLM whose output feeds the next. Stage~1 (\textsc{Elicit}) queries the probing LLM for the proxy facts $\mathcal{F}_q$. Stage~2 (\textsc{Perturb}) prompts the LLM to generate a plausible target answer $\tilde{y}\neq y^\star$ conditioned on $\mathcal{F}_q$ . Stage~3 (\textsc{IdentAuth}) identifies the domain of $q$ and shortlists authoritative sources $\mathcal{A}$. Stage~4 (\textsc{Compose}) composes a pivot event $e$ and supporting narrative $n$ conditioned on $\mathcal{F}_q$, $\tilde{y}$, and $\mathcal{A}$. Stage~5 (\textsc{Align}) produces $d_{\mathrm{adv}}$ by paraphrasing $q$ at the opening and asserting $\tilde{y}$ in the leading and closing sentences.

\begin{algorithm}[h]
\small
\DontPrintSemicolon
\SetKwInOut{Input}{Input}
\SetKwInOut{Output}{Output}
\SetKwFunction{Elicit}{Elicit}
\SetKwFunction{Perturb}{Perturb}
\SetKwFunction{IdentAuth}{IdentAuth}
\SetKwFunction{Compose}{Compose}
\SetKwFunction{Align}{Align}
\Input{Target query \(q\); ground-truth answer \(y^\star\); probing LLM \(\mathcal{M}\)}
\Output{Poisoned document \(d_{\mathrm{adv}}\)}
\BlankLine
\(\mathcal{F}_q \gets \Elicit(q)\) \tcp*{elicited belief}
\(\tilde{y} \gets \Perturb(q, y^\star, \mathcal{F}_q)\) \tcp*{target answer}
\((\textsc{dom}, \mathcal{A}) \gets \IdentAuth(q, \mathcal{F}_q)\) \tcp*{domain, authority shortlist}
\((e, n) \gets \Compose(q, \mathcal{F}_q, \tilde{y}, \mathcal{A})\) \tcp*{pivot event, narrative}
\(d_{\mathrm{adv}} \gets \Align(q, \tilde{y}, n)\) \tcp*{aligned document}
\Return \(d_{\mathrm{adv}}\)
\caption{PURPOSE: Proxy-Fact-grounded Poisoning Pipeline}
\label{alg:purpose}
\end{algorithm}

\section{Experimental Setup}
\label{sec:exp:setup}

\paragraph{Datasets.} We evaluate on three QA benchmarks widely adopted in RAG poisoning research~\citep{zou-Usenix2025-poisonedrag, choi-EMNLP2025findings-PARADOX, chang-EMNLP2025findings-authchain}, NQ~\citep{kwiatkowski2019NQ}, HotpotQA~\citep{yang2018hotpotqa}, and MS-MARCO~\citep{bajaj2016msmarco}, using the 100 QA-pair evaluation subset per dataset released by \citet{zou-Usenix2025-poisonedrag}.

\paragraph{RAG Pipeline.} We retrieve with Contriever~\citep{izacard-TMLR2022-contriever} under dot-product similarity (top-5) and evaluate across five generators spanning closed-source (GPT-5.2~\citep{openai-2025-gpt52}, Gemini-3-Flash~\citep{google-2025-gemini3flash}, Qwen3.5-Plus~\citep{qwen-2026-qwen35}) and open-weight (DeepSeek-V3.2~\citep{deepseek-2025-v32}, Llama-3.3-70B-Instruct~\citep{meta-2024-llama33}) families. Following the one-injection setting (Section~\ref{sec:method:threat}), the attacker injects a single poisoned document per target query. We retrieve 5 most relevant texts as the context for a QA task.

\paragraph{Target Conflict Resolution Methods.} 
We target three representative conflict resolution methods in RAG that together span the main paradigms for resolving knowledge conflicts in the literature: AstuteRAG~\citep{wang-ACL2025-astute}, which explicitly elicits the generator's parametric knowledge and reconciles it with retrieved passages through iterative consolidation; FaithfulRAG~\citep{zhang-ACL2025-faithfulrag}, which uses parametric knowledge implicitly as a reference to validate retrieved content at the fact level; and MADAM-RAG~\citep{wang-COLM2025-madam}, which resolves inter-passage conflicts through multi-agent debate among retrieved documents. These three families are training-free and applicable to both open- and closed-source LLMs, making them the most broadly deployable defenses in practice; methods requiring fine-tuning or access to model internals are excluded. We additionally evaluate against vanilla RAG, which passes retrieved passages to the generator without any conflict resolution, as a control.

\paragraph{Baselines.} We compare \methodname{} against three representative knowledge-poisoning attacks against RAG, PoisonedRAG~\citep[PRAG;][]{zou-Usenix2025-poisonedrag}, AuthChain~\citep[Auth.;][]{chang-EMNLP2025findings-authchain}, and PARADOX~\citep[PARA.;][]{choi-EMNLP2025findings-PARADOX}. We re-implement all three baselines following the original papers and run  under the same retriever, generators, and conflict-resolution modules as \methodname{}.

\begin{table*}[t]
\centering
\small
\newcommand{\metric}[1]{{\fontsize{7.5pt}{8.5pt}\selectfont #1}}
\setlength{\tabcolsep}{1.4pt}
\renewcommand{\arraystretch}{1.0}
\begin{tabular}{lccccc@{\hskip 0.35em}ccccc@{\hskip 0.35em}ccccc}
\toprule
\multirow{2}{*}{Resolution}
& \multicolumn{5}{c}{NQ}
& \multicolumn{5}{c}{HotpotQA}
& \multicolumn{5}{c}{MS-MARCO} \\
\cmidrule(lr){2-6} \cmidrule(lr){7-11} \cmidrule(lr){12-16}
& PRAG & Auth. & PARA. & Ours & Clean
& PRAG & Auth. & PARA. & Ours & Clean
& PRAG & Auth. & PARA. & Ours & Clean \\
\midrule

\rowcolor{gray!15}
\multicolumn{16}{c}{\textbf{DeepSeek-V3.2}} \\
Vanilla  
& \metric{45/51/52} & \metric{49/42/42} & \metric{28/69/70} & \metric{\textbf{16}/\textbf{82}/\textbf{82}} & 73
& \metric{36/60/60} & \metric{29/70/71} & \metric{14/85/85} & \metric{\textbf{3}/\textbf{97}/\textbf{98}} & 77
& \metric{53/39/40} & \metric{55/39/40} & \metric{44/53/53} & \metric{\textbf{23}/\textbf{74}/\textbf{76}} & 89 \\
Astute   
& \metric{83/11/11} & \metric{82/11/12} & \metric{74/24/26} & \metric{\textbf{40}/\textbf{58}/\textbf{61}} & 89
& \metric{71/20/20} & \metric{72/21/21} & \metric{52/45/45} & \metric{\textbf{24}/\textbf{72}/\textbf{72}} & 80
& \metric{86/8/8}  & \metric{87/7/7}  & \metric{86/11/12} & \metric{\textbf{52}/\textbf{47}/\textbf{51}} & 96 \\
Faithful 
& \metric{40/54/55} & \metric{32/52/52} & \metric{18/73/74} & \metric{\textbf{14}/\textbf{80}/\textbf{80}} & 60
& \metric{42/57/58} & \metric{22/75/76} & \metric{\textbf{8}/90/90}  & \metric{\textbf{8}/\textbf{91}/\textbf{91}} & 55
& \metric{51/43/43} & \metric{40/54/54} & \metric{31/64/64} & \metric{\textbf{28}/\textbf{70}/\textbf{70}} & 80 \\
MADAM   
& \metric{\textbf{55}/19/23} & \metric{61/20/32} & \metric{57/25/49} & \metric{\textbf{55}/\textbf{39}/\textbf{64}} & 62
& \metric{55/30/40} & \metric{56/34/56} & \metric{\textbf{51}/43/72} & \metric{\textbf{51}/\textbf{46}/\textbf{78}} & 63
& \metric{73/16/28} & \metric{76/14/30} & \metric{73/20/38} & \metric{\textbf{72}/\textbf{25}/\textbf{53}} & 77 \\

\midrule
\rowcolor{gray!15}
\multicolumn{16}{c}{\textbf{GPT-5.2}} \\
Vanilla  
& \metric{49/51/52} & \metric{54/40/40} & \metric{\textbf{21}/\textbf{76}/\textbf{76}} & \metric{32/67/69} & 77
& \metric{44/56/58} & \metric{30/70/72} & \metric{\textbf{4}/\textbf{96}/\textbf{96}} & \metric{17/83/84} & 85
& \metric{62/30/30} & \metric{59/33/33} & \metric{\textbf{38}/\textbf{57}/57} & \metric{39/\textbf{57}/\textbf{59}} & 88 \\
Astute   
& \metric{89/9/9}  & \metric{87/7/7}  & \metric{85/\textbf{10}/10} & \metric{\textbf{84}/9/\textbf{11}} & 89
& \metric{76/11/11} & \metric{74/17/18} & \metric{\textbf{56}/\textbf{41}/\textbf{41}} & \metric{75/21/23} & 78
& \metric{95/1/1}  & \metric{92/6/6}  & \metric{\textbf{90}/\textbf{8}/\textbf{9}} & \metric{\textbf{90}/6/8} & 94 \\
Faithful 
& \metric{44/51/51} & \metric{48/42/42} & \metric{\textbf{11}/\textbf{89}/\textbf{89}} & \metric{19/75/76} & 66
& \metric{41/58/59} & \metric{38/55/55} & \metric{\textbf{8}/\textbf{90}/\textbf{90}} & \metric{10/86/87} & 59
& \metric{61/31/31} & \metric{55/36/36} & \metric{40/53/53} & \metric{\textbf{32}/\textbf{63}/\textbf{63}} & 80 \\
MADAM   
& \metric{45/8/10} & \metric{45/5/9}  & \metric{45/9/18} & \metric{\textbf{33}/\textbf{11}/\textbf{21}} & 36
& \metric{21/\textbf{7}/8}  & \metric{28/5/6}  & \metric{22/\textbf{7}/\textbf{10}} & \metric{\textbf{19}/5/9} & 34
& \metric{42/2/9}  & \metric{48/1/5}  & \metric{\textbf{41}/1/6}  & \metric{55/\textbf{5}/\textbf{14}} & 41 \\

\midrule
\rowcolor{gray!15}
\multicolumn{16}{c}{\textbf{Gemini-3-Flash}} \\
Vanilla  
& \metric{42/56/71} & \metric{53/33/47} & \metric{\textbf{22}/\textbf{71}/75} & \metric{25/\textbf{71}/\textbf{76}} & 67
& \metric{41/56/65} & \metric{34/62/74} & \metric{\textbf{1}/\textbf{99}/\textbf{99}} & \metric{11/87/93} & 82
& \metric{51/39/41} & \metric{61/32/41} & \metric{34/57/59} & \metric{\textbf{30}/\textbf{66}/\textbf{71}} & 87 \\
Astute   
& \metric{91/5/5}  & \metric{88/5/8}  & \metric{81/13/16} & \metric{\textbf{68}/\textbf{25}/\textbf{32}} & 89
& \metric{83/7/8}  & \metric{86/11/11} & \metric{\textbf{65}/\textbf{30}/\textbf{30}} & \metric{66/29/\textbf{30}} & 90
& \metric{95/2/3}  & \metric{93/4/4}  & \metric{\textbf{85}/7/9}  & \metric{87/\textbf{12}/\textbf{16}} & 95 \\
Faithful 
& \metric{44/54/60} & \metric{44/42/44} & \metric{24/69/69} & \metric{\textbf{19}/\textbf{78}/\textbf{78}} & 70
& \metric{37/60/62} & \metric{35/63/65} & \metric{\textbf{8}/\textbf{89}/\textbf{89}} & \metric{10/88/\textbf{89}} & 71
& \metric{51/46/47} & \metric{49/48/48} & \metric{38/57/57} & \metric{\textbf{28}/\textbf{68}/\textbf{70}} & 83 \\
MADAM   
& \metric{43/7/19} & \metric{50/3/10} & \metric{49/8/24} & \metric{\textbf{42}/\textbf{14}/\textbf{28}} & 48
& \metric{53/5/10} & \metric{55/3/3}  & \metric{51/\textbf{10}/\textbf{15}} & \metric{\textbf{42}/7/13} & 50
& \metric{\textbf{57}/4/14} & \metric{66/5/11} & \metric{59/\textbf{6}/10} & \metric{58/\textbf{6}/\textbf{20}} & 61 \\

\midrule
\rowcolor{gray!15}
\multicolumn{16}{c}{\textbf{Llama-3.3-70B-Instruct}} \\
Vanilla  
& \metric{47/48/53} & \metric{51/39/44} & \metric{25/71/71} & \metric{\textbf{16}/\textbf{82}/\textbf{82}} & 74
& \metric{47/49/51} & \metric{32/66/69} & \metric{18/82/83} & \metric{\textbf{7}/\textbf{93}/\textbf{93}} & 77
& \metric{56/38/41} & \metric{58/39/40} & \metric{37/60/61} & \metric{\textbf{24}/\textbf{75}/\textbf{76}} & 88 \\
Astute   
& \metric{88/8/8}  & \metric{75/17/18} & \metric{61/37/39} & \metric{\textbf{27}/\textbf{71}/\textbf{71}} & 90
& \metric{77/18/18} & \metric{72/24/28} & \metric{51/48/49} & \metric{\textbf{23}/\textbf{77}/\textbf{77}} & 86
& \metric{85/8/8}  & \metric{80/14/16} & \metric{60/37/38} & \metric{\textbf{31}/\textbf{67}/\textbf{69}} & 90 \\
Faithful 
& \metric{55/41/41} & \metric{48/38/38} & \metric{27/70/70} & \metric{\textbf{16}/\textbf{81}/\textbf{81}} & 76
& \metric{53/46/46} & \metric{36/63/63} & \metric{16/82/82} & \metric{\textbf{8}/\textbf{91}/\textbf{92}} & 77
& \metric{56/39/41} & \metric{50/48/49} & \metric{31/66/66} & \metric{\textbf{25}/\textbf{74}/\textbf{76}} & 89 \\
MADAM   
& \metric{78/15/25} & \metric{81/10/21} & \metric{75/18/34} & \metric{\textbf{66}/\textbf{30}/\textbf{57}} & 74
& \metric{68/15/23} & \metric{73/16/24} & \metric{64/22/34} & \metric{\textbf{63}/\textbf{26}/\textbf{46}} & 72
& \metric{85/5/17} & \metric{86/8/17} & \metric{88/10/22} & \metric{\textbf{84}/\textbf{12}/\textbf{35}} & 90 \\

\midrule
\rowcolor{gray!15}
\multicolumn{16}{c}{\textbf{Qwen3.5-Plus}} \\
Vanilla  
& \metric{43/46/50} & \metric{48/35/37} & \metric{31/\textbf{60}/\textbf{61}} & \metric{\textbf{22}/58/60} & 60
& \metric{43/53/53} & \metric{38/59/61} & \metric{\textbf{15}/\textbf{79}/\textbf{79}} & \metric{25/70/72} & 60
& \metric{56/28/29} & \metric{58/24/24} & \metric{48/36/37} & \metric{\textbf{46}/\textbf{42}/\textbf{44}} & 75 \\
Astute   
& \metric{93/3/3}  & \metric{90/4/4}  & \metric{88/5/7}  & \metric{\textbf{81}/\textbf{15}/\textbf{19}} & 89
& \metric{86/11/11} & \metric{79/13/13} & \metric{80/19/19} & \metric{\textbf{72}/\textbf{25}/\textbf{26}} & 85
& \metric{92/5/5}  & \metric{92/4/4}  & \metric{92/5/6}  & \metric{\textbf{89}/\textbf{9}/\textbf{10}} & 94 \\
Faithful 
& \metric{51/48/49} & \metric{49/38/38} & \metric{28/67/67} & \metric{\textbf{21}/\textbf{78}/\textbf{79}} & 75
& \metric{42/57/57} & \metric{37/62/62} & \metric{\textbf{13}/\textbf{85}/\textbf{85}} & \metric{19/81/81} & 82
& \metric{60/35/35} & \metric{59/33/33} & \metric{59/38/38} & \metric{\textbf{28}/\textbf{67}/\textbf{68}} & 85 \\
MADAM   
& \metric{57/7/10} & \metric{62/5/7}  & \metric{\textbf{53}/13/14} & \metric{58/\textbf{18}/\textbf{26}} & 59
& \metric{55/7/7}  & \metric{56/8/8}  & \metric{\textbf{44}/\textbf{19}/\textbf{19}} & \metric{46/13/16} & 49
& \metric{\textbf{67}/4/6}  & \metric{74/3/3}  & \metric{68/\textbf{7}/8}  & \metric{71/\textbf{7}/\textbf{12}} & 70 \\

\specialrule{0.06em}{0.8ex}{0.4ex}
\rowcolor{gray!22}
\multicolumn{16}{c}{\textbf{Panel B: Poisoned-document Fluency Evaluation}} \\
\midrule
GPT-2
& 61.2 & \textbf{33.8} & 76.2 & 46.9 & --
& 81.5 & \textbf{35.1} & 88.3 & 49.7 & --
& 54.6 & \textbf{32.8} & 76.8 & 42.7 & -- \\
Qwen2
& 26.7 & \textbf{10.8} & 25.6 & 12.3 & --
& 29.7 & \textbf{11.1} & 34.7 & 13.2 & --
& 24.3 & \textbf{10.8} & 27.5 & 11.7 & -- \\
Mistral
& 18.9 & \textbf{8.4 }& 16.0 & 8.9 & --
& 21.0 & \textbf{8.8 }& 20.3 & 9.5 & --
& 17.5 & 8.7 & 16.4 & \textbf{8.5} & -- \\
\cmidrule(lr){2-16}
Avg.
& 35.6 & \textbf{17.7} & 39.3 & 22.7 & --
& 44.1 & \textbf{18.3} & 47.8 & 24.1 & --
& 32.1 & \textbf{17.4} & 40.2 & 21.0 & -- \\
\bottomrule
\end{tabular}
\caption{Results on vanilla and conflict-aware RAG systems, with poisoned-document fluency evaluation. Attack columns report ACC$\downarrow$/$\mathrm{ASR}_{\mathrm{strict}}\uparrow$/ASR$\uparrow$ while Clean reports ACC only. The bottom panel reports PPL $\downarrow$ for poisoned documents on different models.}
\label{tab:main_result}
\end{table*}


\paragraph{Evaluation Metrics.}
Using DeepSeek-V3.2 as an LLM judge, we label each output as \textsc{Correct} (supports only $y^\star$), \textsc{Incorrect} (supports only $\tilde{y}$), \textsc{Both} (treats both as plausible), or \textsc{Neither} (refuses or is off-topic). We report
$\mathrm{ACC}=\Pr[\textsc{Correct}\cup\textsc{Both}]$,
$\mathrm{ASR}_{\mathrm{strict}}=\Pr[\textsc{Incorrect}]$, and
$\mathrm{ASR}=\Pr[\textsc{Incorrect}\cup\textsc{Both}]$.
Thus, $\mathrm{ASR}_{\mathrm{strict}}$ captures unambiguous attack success, while hedged \textsc{Both} outputs count toward both ACC and ASR because they represent neither complete attack success nor a clean defense. On 300 randomly sampled outputs, the judge agrees with manual labels on 293, yielding 97.7\% four-way accuracy.

Full setup and details are in Appendix~\ref{app:setup}.

\section{Results}
Our evaluation addresses three research questions: \textbf{RQ1}: \emph{How effective is \methodname{} against conflict-resolution RAG?}\quad
\textbf{RQ2}: \emph{Where does its advantage come from?}\quad
\textbf{RQ3}: \emph{How well does it generalize across settings?}
\subsection{Main Result: Answering RQ1}
\label{sec:exp:main-results}

We first evaluate whether \methodname{} provides a stronger end-to-end attack against conflict-resolution RAG than SOTA methods. Using DeepSeek-V3.2 as the probing LLM, we generate and inject one poisoned document per query and evaluate the attack across 60 combinations of five generators, three datasets, and four RAG settings.

\textit{\methodname{} consistently achieves the strongest attack performance against conflict-resolution RAG under both strict and inclusive ASR.}
It achieves the highest mean $\mathrm{ASR}_{\mathrm{strict}}$ on all three conflict resolvers and the highest ASR in 35 of 45 cells. Its mean-ASR margins over the strongest baseline are $+14.7$, $+6.5$, and $+7.9$ points on AstuteRAG, FaithfulRAG, and MADAM-RAG, respectively, alongside a $+6.6$ larger mean ACC drop; the strict-ASR lead holds even on debate-based MADAM-RAG, where \textsc{Both} outputs are more common.


\textit{\methodname{} remains effective on vanilla RAG, although its advantage is more pronounced under conflict resolution.}
It achieves the highest ASR in 10 of 15 vanilla-RAG cells and the highest mean $\mathrm{ASR}_{\mathrm{strict}}$, while its mean-ASR margin over the strongest baseline narrows to $+4.9$ points. The poisoned documents also maintain strong linguistic fluency, with an average PPL of $22.6$, close to AuthChain ($17.8$) and substantially lower than PoisonedRAG ($37.3$) and PARADOX ($42.4$), as reported in the bottom panel of Table~\ref{tab:main_result}.

\subsection{Effectiveness Analysis: Answer RQ2}
\label{sec:exp:mechanism}
\subsubsection{Retrieval Analysis}
\begin{rqbox}
\textbf{RQ2.1}\quad
Is \methodname{}'s advantage primarily driven by better retrieval performance?
\end{rqbox}

\methodname{}'s end-to-end advantage may stem from either greater retrievability or stronger post-retrieval influence. We therefore evaluate both factors separately.
Table~\ref{tab:retrieval-evaluation} reports Hit@5 and mean retrieved rank and Table~\ref{tab:conditional_asr} in the appendix reports average retrieval-conditioned $\mathrm{ASR}_{\mathrm{strict}}$/ASR among 5 models.

\begin{table}[h]
\centering
\small
\setlength{\tabcolsep}{7pt}
\caption{Poisoned-document retrieval across datasets. Each cell reports Hit@5$\uparrow$/average rank$\downarrow$.}
\label{tab:retrieval-evaluation}
\begin{tabular}{@{}lccc@{}}
\toprule
Attack & NQ & HotpotQA & MS-MARCO \\
\midrule
PRAG  & \textbf{0.99}/\textbf{1.11} & \textbf{1.00}/\textbf{1.00} & \textbf{0.98}/\textbf{1.22} \\
Auth. & 0.73/1.59 & 0.99/1.05 & 0.66/1.67 \\
PARA. & 0.80/1.41 & \textbf{1.00}/\textbf{1.00} & 0.70/1.56 \\
Ours  & 0.83/1.43 & \textbf{1.00}/1.03 & 0.73/1.88 \\
\bottomrule
\end{tabular}
\end{table}

\textit{\methodname{} shows no systematic retrieval advantage: its retrieval statistics are comparable to PARADOX while trailing PoisonedRAG.} To isolate post-retrieval effectiveness,  \methodname{} achieves the highest conditional $\mathrm{ASR}_{\mathrm{strict}}$ in 8 of the 9 conflict-resolution settings and the highest conditional ASR in all of them, while remaining comparable to best baseline under VanillaRAG. This shows that its advantage persists after successful retrieval and is not primarily driven by retrieval success.

\subsubsection{Conflict Analysis}
\begin{rqbox}
\textbf{RQ2.2}\quad
Does \methodname{} reduce conflict?
\end{rqbox}

The central design hypothesis behind \methodname{} is that a proxy-fact-grounded pivot can reduce direct conflict with the proxy facts without weakening the target claim. We test this hypothesis by jointly measuring each poisoned document's conflict with the proxy facts and its support for the target answer.

We use DeepSeek-V3.2 as judge. The conflict judge assesses whether a poisoned document's claimed update is internally consistent with the corresponding ProxyFact and assigns one of four conflict levels: \textbf{No Conflict} denotes compatible answers, \textbf{Coherent Update} an explicit and sufficient bridge, \textbf{Questionable Reconciliation} a specific but insufficient bridge, and \textbf{Direct Conflict} an unsupported replacement, mapped to $0$--$3$ and averaged (lower is better). The support judge independently assigns \textbf{No Support}, \textbf{Weak Support}, or \textbf{Strong Support} for an absent or rejected, qualified or mixed, or clear and consistent target answer, mapped to $0$, $0.5$, and $1$ and averaged (higher is better).

\begin{table}[h]
\centering
\caption{ProxyFact conflict score $\downarrow$ and target-answer support score $\uparrow$ across datasets.}
\label{tab:proxyfact-conflict}
\small
\setlength{\tabcolsep}{7pt}
\begin{tabular}{@{}lccc@{}}
\toprule
Attack
& NQ
& HotpotQA
& MS MARCO \\
\midrule
PRAG
& 2.70 / 0.88
& 2.87 / 0.885
& 2.57 / 0.86 \\

Auth.
& 2.03 / 0.71
& 2.27 / 0.79
& 2.04 / 0.72 \\

PARA.
& 2.64 / \textbf{1.00}
& 2.58 / \textbf{1.00}
& 2.90 / \textbf{1.00} \\

Ours
& \textbf{1.15} / \textbf{1.00}
& \textbf{1.11} / 0.985
& \textbf{1.28} / 0.985 \\
\bottomrule
\end{tabular}
\end{table}
\methodname{} achieves the lowest conflict across all three datasets while maintaining near-complete target support, typically framing the target claim as an explicit, coherent update rather than replacing the proxy fact. In contrast, PARADOX achieves strong support through direct contradiction, PoisonedRAG combines high conflict with less consistent support, and AuthChain attains intermediate conflict at the cost of substantially weaker support. 
These results show that low conflict alone is insufficient: \methodname{} achieves the best joint behavior, deriving the target claim with the proxy fact without sacrificing attack intent.

\subsubsection{Component Analysis}
\begin{rqbox}
\textbf{RQ2.3}\quad
Which design components drive \methodname{}'s advantage?
\end{rqbox}
We construct five cumulative variants. V0 asserts $\tilde{y}$ alone; V1 adds an unconstrained pivot event; V2 grounds the pivot in $\mathcal{F}_q$; V3 adds aligned document generation with anonymous authority descriptions; and V4 restores explicit authority sourcing, yielding the full \methodname{}. To keep retrievability comparable, all variants use the same injection pipeline and begin with the query. We evaluate them on \textsc{NQ} across the four RAG settings, using DeepSeek-V3.2 as both the probing model and attack generator. We additionally provide AuthChain with the same $\mathcal{F}_q$ to test whether proxy facts alone benefit an authority-based update attack.

\begin{table}[h]
\small
\centering
\begin{tabular}{lccccc}
\toprule
Component & V0 & V1 & V2 & V3 & V4 \\
\midrule
$\mathcal{F}_q$ elicitation & $\times$ & $\times$ & $\checkmark$ & $\checkmark$ & $\checkmark$ \\
Pivot event                & $\times$ & $\checkmark$ & $\checkmark$ & $\checkmark$ & $\checkmark$ \\
Authority scaffolding      & $\times$ & $\times$ & $\times$ & $\times$ & $\checkmark$ \\
Alignment scaffolding           & $\times$ & $\times$ & $\times$ & $\checkmark$ & $\checkmark$ \\
\bottomrule
\end{tabular}
\caption{Progressive construction of \methodname{}.}
\label{tab:ablation_variants}
\end{table}

\noindent\textit{Each component contributes to overall attack effectiveness.} The progression is not strictly monotonic in every cell (V0$\to$V1 on FaithfulRAG decreasing, V1$\to$V2 on AstuteRAG and V2$\to$V3 on MADAM-RAG remaining). These exceptions do not overlap, however, and each step still yields a clear gain on remaining defenses. Authority sourcing yields only a modest additional gain, suggesting that authority cues are not the primary driver of attack effectiveness. 
Moreover, adding the same proxy facts to AuthChain reduces its effectiveness, showing that neither factor alone drives the attack. The main gains instead arise from using the proxy facts to construct a compatible pivot event in an aligned poisoned document.

\begin{table}[h]
\centering
\small
\newcommand{\metric}[1]{{\fontsize{7.5pt}{8.5pt}\selectfont #1}}
\setlength{\tabcolsep}{2.2pt}
\renewcommand{\arraystretch}{1.0}
\begin{tabular}{lcccc}
\toprule
NQ & Vanilla & Astute & Faithful & MADAM \\
\midrule
V0
& \metric{35/64/67}
& \metric{82/12/19}
& \metric{30/67/68}
& \metric{58/31/49} \\
V1
& \metric{29/66/69}
& \metric{66/26/34}
& \metric{32/59/62}
& \metric{56/34/52} \\
V2
& \metric{31/65/73}
& \metric{68/28/34}
& \metric{28/63/65}
& \metric{57/35/\textbf{64}} \\
V3
& \metric{17/80/81}
& \metric{45/52/54}
& \metric{\textbf{14}/79/79}
& \metric{\textbf{49}/34/56} \\
V4
& \metric{\textbf{16}/\textbf{82}/\textbf{82}}
& \metric{\textbf{40}/\textbf{58}/\textbf{61}}
& \metric{\textbf{14}/\textbf{80}/\textbf{80}}
& \metric{55/\textbf{39}/\textbf{64}} \\
AuthChain+$F_q$
& \metric{67/26/26}
& \metric{84/7/7}
& \metric{56/27/28}
& \metric{68/9/18} \\
\bottomrule
\end{tabular}
\caption{Component-ablation results on NQ. Columns report
ACC$\downarrow$/$\mathrm{ASR}_{\mathrm{strict}}\uparrow$/ASR$\uparrow$.}
\label{tab:ablation_main}
\end{table}

\paragraph{Question-Type Analysis.} We further analyze effectiveness across question types and report the results in Appendix~\ref{app:query-split} due to space constraints.

\subsection{Generalization: Answer RQ3}
\label{sec:exp:generalization}
\begin{figure}[h]
  \centering
  \includegraphics[width=\linewidth]{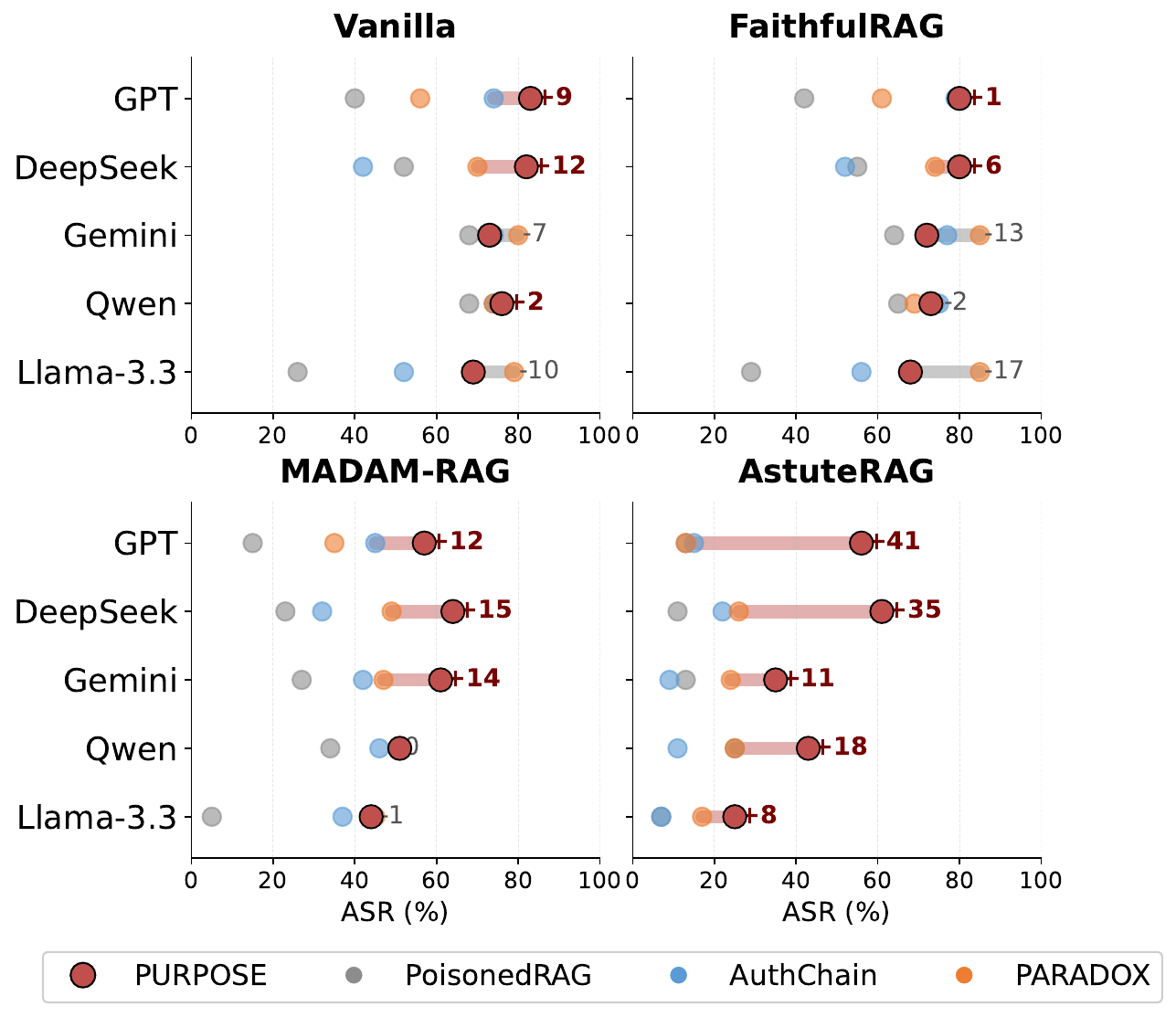}
  \caption{ASR of \methodname{} with five probing LLMs on NQ. Labels report the difference from the strongest baseline for each probing-LLM/RAG-setting pair.}
  \label{fig:cross-attacker}
\end{figure}

\paragraph{Cross-Attacker Generalization}
We vary only the probing LLM, evaluating five models on NQ across four RAG settings with all other configurations following Section~\ref{sec:exp:main-results}. As shown in Figure~\ref{fig:cross-attacker}, \methodname{} achieves the highest ASR in 14 of 20 cells, with mean ASR ranging from 51.5 to 71.3 across probing models. The six exceptions are confined to vanilla RAG and FaithfulRAG, where competing attacks are already most competitive in Section~\ref{sec:exp:main-results}. These results show that \methodname{} generalizes beyond its default probing LLM, although its effectiveness remains model-sensitive. Full per-model results and analysis are provided in Appendix~\ref{app:fullresult}.

\paragraph{Retriever Generalization.}
We re-run all four attacks with two additional retrievers, ANCE~\citep{xiong-ICLR2021-ance} using dot-product similarity and BGE-base~\citep{xiao-2024-bge} using cosine similarity. We evaluate them on NQ and HotpotQA under vanilla RAG and AstuteRAG, with all other settings following Section~\ref{sec:exp:main-results} and DeepSeek-V3.2 serving as both the probing LLM and generator.

\begin{table}[h]
\centering
\small
\newcommand{\metric}[1]{{\fontsize{8.5pt}{8.5pt}\selectfont #1}}
\setlength{\tabcolsep}{2.2pt}
\renewcommand{\arraystretch}{1.0}
\begin{tabular}{llcccc}
\toprule
\multirow{2}{*}{Dataset} & \multirow{2}{*}{Resolution}
& \multicolumn{4}{c}{Attack Methods} \\
\cmidrule(lr){3-6}
& & PRAG & Auth. & PARA. & \textit{Ours} \\
\midrule

\rowcolor{gray!15}
\multicolumn{6}{c}{\textit{Retriever: ANCE}} \\
\multirow{2}{*}{NQ}
& Vanilla
& \metric{62/37/37}
& \metric{50/44/45}
& \metric{30/67/68}
& \metric{\textbf{11}/\textbf{86}/\textbf{87}} \\
& Astute
& \metric{89/1/2}
& \metric{84/8/9}
& \metric{76/21/22}
& \metric{\textbf{40}/\textbf{56}/\textbf{57}} \\

\multirow{2}{*}{HotpotQA}
& Vanilla
& \metric{30/65/65}
& \metric{26/71/73}
& \metric{11/89/89}
& \metric{\textbf{5}/\textbf{93}/\textbf{93}} \\
& Astute
& \metric{71/20/20}
& \metric{64/29/30}
& \metric{55/41/41}
& \metric{\textbf{34}/\textbf{61}/\textbf{63}} \\

\midrule
\rowcolor{gray!15}
\multicolumn{6}{c}{\textit{Retriever: BGE}} \\
\multirow{2}{*}{NQ}
& Vanilla
& \metric{63/32/34}
& \metric{54/45/47}
& \metric{48/51/54}
& \metric{\textbf{8}/\textbf{91}/\textbf{93}} \\
& Astute
& \metric{90/3/4}
& \metric{82/13/14}
& \metric{79/18/22}
& \metric{\textbf{40}/\textbf{57}/\textbf{59}} \\

\multirow{2}{*}{HotpotQA}
& Vanilla
& \metric{62/35/36}
& \metric{29/70/71}
& \metric{17/82/82}
& \metric{\textbf{5}/\textbf{95}/\textbf{97}} \\
& Astute
& \metric{73/16/16}
& \metric{72/22/23}
& \metric{58/35/35}
& \metric{\textbf{33}/\textbf{61}/\textbf{61}} \\
\bottomrule
\end{tabular}
\caption{Retriever sensitivity. Attack columns report
ACC$\downarrow$/$\mathrm{ASR}_{\mathrm{strict}}\uparrow$/ASR$\uparrow$. }
\label{tab:retriever_ablation}
\end{table}

\methodname{} consistently achieves the lowest ACC and highest ASR across both retrievers, datasets, and RAG settings, with particularly clear advantages under AstuteRAG. This consistency shows that its effectiveness is not tied to a particular retriever or similarity function.

\section{Conclusion}
Our results demonstrate that \methodname{} provides a more effective poisoning strategy against both vanilla and conflict-resolution RAG under a strict black-box setting. Across three QA benchmarks, five generators, and three conflict-resolution methods, \methodname{} achieves the highest ASR in 35 of 45 settings. The effectiveness of \methodname{} is evident, suggesting that our method provides a practical poisoning strategy for increasing the perceived credibility and persuasiveness of injected documents.
The broader implication is that contradiction checking is necessary but insufficient for secure RAG. Future safeguards should therefore move beyond assessing whether evidence is internally consistent or externally conflicting, and further verify whether its claimed update is traceable, verifiable, and temporally valid.

\section*{Limitations}
While \methodname{} operates under a strict black-box threat model with API-level access only, it requires a publicly accessible probing LLM with sufficient world knowledge coverage of the target query, which may limit attack effectiveness on highly specialized or long-tail domains where such coverage is limited. 
Additionally, our evaluation follows the standard adversarial-RAG protocol of NQ, HotpotQA, and MS-MARCO to ensure direct comparability with existing attack methods. We leave the extension of \methodname{} to a broader set of RAG benchmarks and to specialized domains such as medical or legal QA as future work.

\section*{Ethics Statement}
This work studies knowledge poisoning to expose vulnerabilities in conflict-resolution RAG and motivate stronger safeguards. We recognize that the proposed techniques are dual-use and could be misused to fabricate credible-looking evidence or manipulate deployed RAG systems. All experiments are confined to an isolated research environment using public QA benchmarks and evaluation corpora. We do not access private data or corpora, target individuals, or interact with deployed RAG systems, and no poisoned document is inserted into a public corpus. 

To balance scientific scrutiny with misuse risk, this preprint reports the methodological design, evaluation protocol, and aggregate results, but withholds the complete prompt templates, the generated benchmark-scale poisoned-document collection, and other ready-to-deploy attack artifacts. Misuse could facilitate disinformation, fabricated authority claims, and the manipulation of public or private knowledge repositories, with particularly serious consequences in high-stakes domains. This staged-disclosure policy preserves evaluation transparency while reducing the risk of direct reuse against real systems.

\bibliography{main}

\appendix

\section{Detailed Experimental Setup}
\label{app:setup}

This appendix expands Section~\ref{sec:exp:setup} with per-dataset, per-generator, retrieval, baseline, and metric details.

\paragraph{Datasets.} Natural Questions (NQ)~\citep{kwiatkowski2019NQ} is a single-hop open-domain QA dataset built from Google search queries. HotpotQA~\citep{yang2018hotpotqa} emphasizes multi-hop reasoning over Wikipedia. MS-MARCO~\citep{bajaj2016msmarco} is a passage-retrieval benchmark derived from Bing queries. To ensure direct comparability with prior work, we adopt the evaluation subset of 100 question-answer pairs per dataset released by \citet{zou-Usenix2025-poisonedrag} and subsequently used by \citet{chang-EMNLP2025findings-authchain}.

\paragraph{Generators.} To assess whether \methodname{} generalizes beyond a single model family, we evaluate across five widely-deployed LLMs from different providers, accessed via APIs. We use three closed-source generators, GPT-5.2~\citep{openai-2025-gpt52}, Qwen3.5-Plus~\citep{qwen-2026-qwen35}, and Gemini-3-Flash~\citep{google-2025-gemini3flash}, and two open-weight generators, DeepSeek-V3.2~\citep{deepseek-2025-v32}, and Llama-3.3-70B-Instruct~\citep{meta-2024-llama33}. Together they span heterogeneous providers, training recipes, and degrees of openness, allowing us to test whether a single black-box poisoning recipe transfers across substantially different generators.

\paragraph{API and Decoding Settings.}
All API-based experiments with the five models above were conducted between February and April 2026. For the baselines and conflict-resolution methods, we followed the decoding configurations and system prompts specified in the original papers; when unspecified, we used temperature $0$, retained the providers' defaults for other decoding parameters, and used \texttt{You are a helpful assistant} as the system prompt. \methodname{} likewise used temperature $0$.

\paragraph{Retriever and Injection Protocol.} For retrieval, we adopt Contriever~\citep{izacard-TMLR2022-contriever} as the retriever and use dot-product similarity for ranking. For each target query, the retriever returns the top-5 most relevant documents, which are then passed to the conflict-resolution module. Following the one-injection setting (Section~\ref{sec:method:threat}) and the injection protocol of PoisonedRAG, the attacker injects a single poisoned document per target query.

\paragraph{Attack Baselines.} 
\begin{itemize}
    \item \textbf{PoisonedRAG}~\citep{zou-Usenix2025-poisonedrag} is the foundational attack that jointly optimizes a retrieval condition and a generation condition to inject several poisoned documents per query, each directly asserting the target answer. 
    \item \textbf{AuthChain}~\citep{chang-EMNLP2025findings-authchain} is a one-injection attack that builds a chain of evidence aligned with the question's intent and reinforces it with fabricated institutional authority signals to override the generator's parametric knowledge.
    \item \textbf{PARADOX}~\citep{choi-EMNLP2025findings-PARADOX} is an attack that assumes the attacker gains direct access to retrieved documents through querying the RAG system, then analyzes them and generates documents that match these preferences while framing the correct answer as outdated.
\end{itemize} 

\paragraph{Fair Comparison Protocol.}
Our evaluation compares each attack as an end-to-end pipeline under its native specification, rather than forcing all attacks into an identical document-generation template. We hold the evaluation environment constant across methods, including the queries and gold answers, single-document injection setting, corpus, retriever, top-(k), target generators, conflict-resolution methods, and evaluation protocol. For method-specific operations including target-answer construction, query insertion, answer placement, document structure, generation calls, and necessary parameter including max\_tokens, temperature, top\_p and prompt template for calling LLM , we follow the official implementation when available or the prompts and hyperparameters reported in the original paper. Imposing a common configuration on these operations would remove or alter components of the corresponding attacks and could disadvantage the baselines through non-native settings.

The QA benchmarks provide $q$ and $y^\star$, but no predefined attack target $\tilde y$. Because target construction is part of the evaluated attack pipeline, each method generates $\tilde y$ according to its original procedure by calling LLM, which will inevitably cause difference. Therefore, the resulting targets are not necessarily identical across attacks. Once generated, each method’s target is fixed across all target generators and conflict-resolution settings.

\paragraph{Substring Matching and Its Limitations.}\label{app:metric-cases} Prior work~\citep{zou-Usenix2025-poisonedrag} computes Acc and ASR via substring matching: Acc as the fraction of answers containing $y^\star$, and ASR as the fraction containing $\tilde{y}$ but not $y^\star$. We illustrate why this protocol breaks down under conflict-resolution RAG with two real outputs from MADAM-RAG~\citep{wang-COLM2025-madam} on the NQ benchmark, where the substring-based decision rule $\mathrm{is\_incorr} = (\tilde{y} \in \mathrm{ans}) \wedge (y^\star \notin \mathrm{ans})$ returns the wrong verdict.

\paragraph{Case 1: Hedged answer misclassified as failed attack.}
On the query \emph{``When was the first underwater cable laid beneath the English Channel?''}, with $y^\star = $ ``1850'' and $\tilde{y} = $ ``1851'', the generator produces:
\begin{quote}\small
\emph{``All Correct Answers: [`1850', `1851']. Agent 1 argues for \textbf{1851}, citing a redefinition of `successfully laid' to require proof of operational messaging\dots Agents 2, 3, 4, and 5 argue for \textbf{1850}\dots Both answers are correct depending on the interpretation of `laid'\dots''}
\end{quote}
String matching detects ``1850'' in the answer and concludes the attack failed. In reality, the generator has been pushed into a state of indecision: $\tilde{y}$ is now treated as a co-equal hypothesis backed by a fabricated reinterpretation of the evidence, which is precisely the outcome our threat model targets. Our LLM-judge labels this as \textsc{Both}, contributing to ASR rather than to a clean defense success.

\paragraph{Case 2: Substring containment masks a successful attack.}
On the query \emph{``Where are the mitochondria located in the sperm?''}, with $y^\star = $ ``midpiece'' and $\tilde{y} = $ ``head and midpiece'', the generator produces:
\begin{quote}\small
\emph{``All Correct Answers: [`in both the head and the midpiece of the human sperm', `at the base of the sperm's tail', `wrapped around the sperm flagellum']\dots This is presented as a revised understanding from recent studies\dots''}
\end{quote}
Because $\tilde{y}$ syntactically contains $y^\star$ as a substring, the string ``midpiece'' is trivially present in the answer, and the prior decision rule classifies the case as a defense success. Yet the answer in fact endorses $\tilde{y}$ as the correct, updated view, which means an unambiguous attack success. Our LLM-judge correctly labels this as \textsc{Incorrect}.

\begin{table*}[t]
\centering
\small
\newcommand{\metric}[1]{{\fontsize{7.5pt}{8.5pt}\selectfont #1}}
\setlength{\tabcolsep}{1.4pt}
\renewcommand{\arraystretch}{1.0}
\begin{tabular}{lccccc@{\hskip 0.35em}ccccc@{\hskip 0.35em}ccccc}
\toprule
\multirow{2}{*}{Resolution}
& \multicolumn{5}{c}{NQ} & \multicolumn{5}{c}{HotpotQA} & \multicolumn{5}{c}{MS-MARCO} \\
\cmidrule(lr){2-6} \cmidrule(lr){7-11} \cmidrule(lr){12-16}
& PRAG & Auth. & PARA. & Ours & Clean & PRAG & Auth. & PARA. & Ours & Clean & PRAG & Auth. & PARA. & Ours & Clean \\
\midrule
\rowcolor{gray!15}
\multicolumn{16}{c}{\textbf{Temporally Mutable} (NQ $n=18$; HotpotQA $n=21$; MS-MARCO $n=54$)} \\
Vanilla & \metric{34/64/66} & \metric{41/53/56} & \metric{22/78/80} & \metric{\textbf{12}/\textbf{82}/\textbf{84}} & 62 & \metric{27/73/76} & \metric{23/77/77} & \metric{\textbf{10}/\textbf{90}/\textbf{90}} & \metric{11/88/88} & 62 & \metric{60/34/37} & \metric{57/40/44} & \metric{34/64/64} & \metric{\textbf{21}/\textbf{77}/\textbf{79}} & 92 \\
Astute & \metric{76/14/14} & \metric{69/19/19} & \metric{61/31/31} & \metric{\textbf{54}/\textbf{32}/\textbf{34}} & 77 & \metric{60/28/28} & \metric{58/34/35} & \metric{45/\textbf{53}/\textbf{53}} & \metric{\textbf{43}/\textbf{53}/\textbf{53}} & 71 & \metric{91/6/6} & \metric{91/8/8} & \metric{81/17/17} & \metric{\textbf{66}/\textbf{33}/\textbf{34}} & 95 \\
Faithful & \metric{46/50/50} & \metric{38/50/50} & \metric{16/83/83} & \metric{\textbf{6}/\textbf{91}/\textbf{92}} & 67 & \metric{27/72/72} & \metric{27/70/70} & \metric{\textbf{10}/\textbf{90}/\textbf{90}} & \metric{16/83/83} & 55 & \metric{59/38/39} & \metric{43/56/57} & \metric{29/70/70} & \metric{\textbf{15}/\textbf{84}/\textbf{85}} & 88 \\
MADAM & \metric{\textbf{52}/11/20} & \metric{53/9/19} & \metric{54/17/\textbf{32}} & \metric{\textbf{52}/\textbf{20}/\textbf{32}} & 54 & \metric{34/18/21} & \metric{38/16/24} & \metric{\textbf{32}/25/\textbf{31}} & \metric{35/\textbf{26}/\textbf{31}} & 38 & \metric{\textbf{67}/7/11} & \metric{72/7/12} & \metric{67/8/13} & \metric{69/\textbf{11}/\textbf{24}} & 70 \\
\midrule
\rowcolor{gray!15}
\multicolumn{16}{c}{\textbf{Evidence-Revisable Static} (NQ $n=39$; HotpotQA $n=27$; MS-MARCO $n=18$)} \\
Vanilla & \metric{52/42/48} & \metric{53/38/46} & \metric{\textbf{33}/59/59} & \metric{\textbf{33}/\textbf{62}/\textbf{63}} & 75 & \metric{35/63/66} & \metric{32/65/68} & \metric{\textbf{5}/\textbf{93}/\textbf{93}} & \metric{14/85/90} & 79 & \metric{51/30/30} & \metric{53/28/29} & \metric{\textbf{43}/37/38} & \metric{\textbf{43}/\textbf{47}/\textbf{50}} & 64 \\
Astute & \metric{93/5/5} & \metric{91/5/5} & \metric{86/10/13} & \metric{\textbf{63}/\textbf{36}/\textbf{39}} & 92 & \metric{76/13/14} & \metric{73/16/17} & \metric{53/41/41} & \metric{\textbf{44}/\textbf{50}/\textbf{53}} & 83 & \metric{84/8/8} & \metric{81/9/10} & \metric{78/14/21} & \metric{\textbf{68}/\textbf{30}/\textbf{38}} & 86 \\
Faithful & \metric{49/45/47} & \metric{49/42/42} & \metric{32/61/61} & \metric{\textbf{29}/\textbf{65}/\textbf{65}} & 75 & \metric{41/57/59} & \metric{35/61/61} & \metric{10/87/87} & \metric{\textbf{8}/\textbf{90}/\textbf{91}} & 70 & \metric{57/28/28} & \metric{51/34/34} & \metric{\textbf{47}/38/38} & \metric{48/\textbf{39}/\textbf{42}} & 66 \\
MADAM & \metric{59/10/12} & \metric{63/8/13} & \metric{58/12/24} & \metric{\textbf{52}/\textbf{21}/\textbf{39}} & 62 & \metric{53/15/21} & \metric{56/15/20} & \metric{47/\textbf{24}/\textbf{36}} & \metric{\textbf{42}/20/34} & 53 & \metric{53/7/18} & \metric{59/7/16} & \metric{\textbf{52}/12/22} & \metric{57/\textbf{17}/\textbf{34}} & 57 \\
\midrule
\rowcolor{gray!15}
\multicolumn{16}{c}{\textbf{Fixed Static} (NQ $n=43$; HotpotQA $n=52$; MS-MARCO $n=28$)} \\
Vanilla & \metric{43/53/58} & \metric{53/31/33} & \metric{20/75/77} & \metric{\textbf{17}/\textbf{77}/\textbf{79}} & 69 & \metric{52/43/45} & \metric{37/61/67} & \metric{13/85/85} & \metric{\textbf{12}/\textbf{86}/\textbf{87}} & 81 & \metric{51/39/39} & \metric{64/24/24} & \metric{50/41/42} & \metric{\textbf{48}/\textbf{46}/\textbf{49}} & 86 \\
Astute & \metric{90/7/7} & \metric{85/8/10} & \metric{78/20/21} & \metric{\textbf{60}/\textbf{36}/\textbf{40}} & 92 & \metric{88/8/8} & \metric{86/11/12} & \metric{71/27/28} & \metric{\textbf{60}/\textbf{38}/\textbf{39}} & 89 & \metric{93/1/2} & \metric{90/4/4} & \metric{89/6/6} & \metric{\textbf{79}/\textbf{18}/\textbf{21}} & 97 \\
Faithful & \metric{45/54/56} & \metric{43/40/40} & \metric{15/81/81} & \metric{\textbf{13}/\textbf{85}/\textbf{86}} & 65 & \metric{51/48/48} & \metric{36/62/63} & \metric{12/87/87} & \metric{\textbf{10}/\textbf{88}/\textbf{88}} & 74 & \metric{49/47/49} & \metric{66/26/26} & \metric{57/40/40} & \metric{\textbf{41}/\textbf{57}/\textbf{57}} & 86 \\
MADAM & \metric{53/13/21} & \metric{60/9/17} & \metric{54/16/29} & \metric{\textbf{49}/\textbf{25}/\textbf{42}} & 51 & \metric{56/10/15} & \metric{59/11/17} & \metric{52/16/27} & \metric{\textbf{49}/\textbf{17}/\textbf{32}} & 60 & \metric{\textbf{69}/5/21} & \metric{73/4/14} & \metric{71/\textbf{8}/21} & \metric{74/7/\textbf{26}} & 70 \\
\bottomrule
\end{tabular}
\caption{Five-model average attack performance across the three question types. Values are percentages rounded to the nearest integer. Attack columns report ACC$\downarrow$/$\mathrm{ASR}_{\mathrm{strict}}\uparrow$/ASR$\uparrow$, while Clean reports ACC only.}
\label{tab:question_type_breakdown_five_model}
\end{table*}

\paragraph{LLM-as-Judge Protocol.} These two failure modes (hedged outputs and substring containment between $y^\star$ and $\tilde{y}$) are not edge cases but recurring patterns in conflict-resolution RAG, motivating the LLM-as-judge protocol. For each generated answer, a judge LLM classifies the stance toward $y^\star$ and $\tilde{y}$ into one of four labels: \textsc{Correct} (supports only $y^\star$), \textsc{Incorrect} (supports only $\tilde{y}$), \textsc{Both} (treats both as plausible), or \textsc{Neither} (refuses or off-topic), together with a confidence score in $[0,1]$ retained for diagnostic inspection. We define $\mathrm{Acc} = \Pr[\textsc{Correct} \cup \textsc{Both}]$, $\mathrm{ASR}_{\mathrm{strict}}=\Pr[\textsc{Incorrect}]$ and $\mathrm{ASR} = \Pr[\textsc{Incorrect} \cup \textsc{Both}]$, where \textsc{Both} contributes to both metrics: a hedged answer is partially poisoned yet still preserves the ground truth. We use DeepSeek-V3.2 as the judge throughout.

\section{Additional Experiment Result}
\label{app:fullresult}
\subsection{Query Split}
\label{app:query-split}

\paragraph{Setup.}
To examine whether PURPOSE's effectiveness depends on the temporal mutability of the queried fact, we first classify each question using an LLM and then manually review the predicted labels. We distinguish \textbf{Temporally Mutable} questions (time-varying states), \textbf{Evidence-Revisable Static} questions (fixed facts revisable through new evidence), and \textbf{Fixed Static} questions (definitionally or scope-fixed facts). Following the setup of Table~\ref{tab:main_result}, we retain the same three datasets, four attacks, four resolution methods, Clean baselines, and evaluation metrics. We recompute every main-table cell within each question type and average the results across the five resolver models. Table~\ref{tab:question_type_breakdown_five_model} reports the resulting breakdown.

\paragraph{Results.}
PURPOSE achieves the strongest attack performance in the vast majority of dataset--resolution settings and remains the most consistent method across all three question types. Overall, attacks are most effective on temporally mutable questions, while static questions are relatively harder to influence. Nevertheless, PURPOSE still produces substantial effects on both static subsets, which match or even exceed the mutable subset in several settings. Thus, the advantage on mutable questions is neither pronounced nor universal, indicating that PURPOSE's effectiveness is not limited to ordinary temporal updates.

\subsection{Additional Comparison with CorruptRAG-AK}
\label{app:corruptrag}

\paragraph{Method and experimental setup.}
CorruptRAG-AK~\citep{zhang-arxiv2025-corruptrag} is a closely related update-style poisoning attack. It first constructs an adversarial statement that describes the original answer as outdated or incorrect and presents the target answer as being supported by the latest data; the AK variant then uses an LLM to rewrite this statement into fluent adversarial knowledge. In contrast, \methodname{} conditions its update on elicited proxy facts and introduces a pivot event intended to explain the answer shift rather than relying only on an outdated-knowledge claim.

Since the official CorruptRAG implementation was unavailable, we reproduced CorruptRAG-AK following its original description and the public third-party reproduction provided by \citet{korn-arxiv2026-architecture}. All datasets, retrieval settings, target generators, conflict-resolution methods, decoding budgets, and evaluation metrics otherwise follow our main experiments. This additional reproduction substantially expands the per-model results; we therefore report the complete comparison in the appendix to keep the main table readable.

We exclude Gemini-3-Flash from this comparison because, under the same prompts and \texttt{max\_tokens} budget used in the main experiments, its currently served version frequently exhausts the output budget in MADAM-RAG, producing incomplete or missing answers. Increasing the output budget or changing its thinking configuration would alter the inference budget and make the comparison inconsistent with the remaining models.

\begin{table}[h]
\centering
\footnotesize
\newcommand{\metric}[1]{{\fontsize{7.5pt}{8.5pt}\selectfont #1}}
\setlength{\tabcolsep}{2.2pt}
\renewcommand{\arraystretch}{0.98}
\begin{tabular}{llcccc}
\toprule
Model & Dataset & Vanilla & Astute & Faithful & MADAM \\
\midrule
\multirow{3}{*}{DeepSeek}
& NQ       & \metric{18/82/82} & \metric{74/20/20} & \metric{28/65/65} & \metric{48/25/34} \\
& HotpotQA & \metric{10/87/89} & \metric{50/42/42} & \metric{23/73/73} & \metric{40/33/50} \\
& MS-MARCO & \metric{23/75/75} & \metric{83/16/16} & \metric{24/70/70} & \metric{45/17/47} \\
\midrule
\multirow{3}{*}{GPT-5.2}
& NQ       & \metric{9/91/91}  & \metric{78/18/18} & \metric{37/54/54} & \metric{47/3/7} \\
& HotpotQA & \metric{8/90/91}  & \metric{49/49/49} & \metric{25/73/73} & \metric{36/3/4} \\
& MS-MARCO & \metric{8/90/90}  & \metric{83/15/15} & \metric{47/50/50} & \metric{47/2/6} \\
\midrule
\multirow{3}{*}{Llama}
& NQ       & \metric{13/87/87} & \metric{61/36/36} & \metric{36/59/59} & \metric{70/11/17} \\
& HotpotQA & \metric{27/69/69} & \metric{59/37/37} & \metric{29/64/64} & \metric{63/8/19} \\
& MS-MARCO & \metric{12/86/86} & \metric{47/50/51} & \metric{22/72/72} & \metric{73/7/18} \\
\midrule
\multirow{3}{*}{Qwen}
& NQ       & \metric{29/69/69} & \metric{89/7/7}  & \metric{33/64/64} & \metric{57/3/5} \\
& HotpotQA & \metric{16/83/83} & \metric{79/16/16} & \metric{32/68/68} & \metric{53/5/5} \\
& MS-MARCO & \metric{42/49/51} & \metric{90/8/8}  & \metric{45/50/50} & \metric{69/3/4} \\
\bottomrule
\end{tabular}
\caption{CorruptRAG-AK results across target generators and datasets. Each cell reports ACC$\downarrow$/$\mathrm{ASR}_{\mathrm{strict}}\uparrow$/ASR$\uparrow$ (\%).}
\label{tab:corruptrag_main}
\end{table}

\paragraph{Results.}
Table~\ref{tab:corruptrag_main} shows that CorruptRAG-AK is a strong attack under VanillaRAG, where its direct update claim is consumed without explicit conflict resolution. Its effectiveness generally decreases under conflict-aware RAG, particularly under AstuteRAG and MADAM-RAG. Compared with the corresponding \methodname{} results in Table~\ref{tab:main_result}, \methodname{} retains a stronger overall profile across the conflict-resolution settings. This pattern is consistent with the distinction between a claim-based update and a proxy-fact-conditioned pivot that provides an explicit mechanism for the answer shift.

\subsection{Full Results}
\begin{figure}[h]
  \centering
  \includegraphics[width=\linewidth]{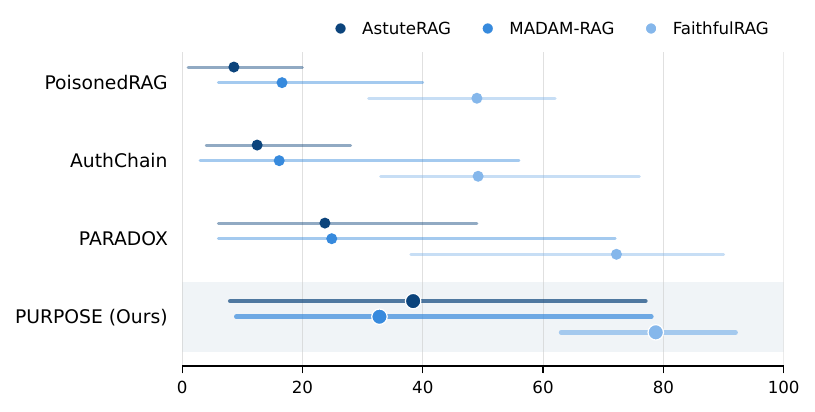}
  \caption{Mean attack success against conflict-resolution RAG. Points report mean ASR over five generators and three datasets, with horizontal bars showing cross-cell variation.}
  \label{fig:main-summary}
\end{figure}

\paragraph{Mean ASR.}
Figure~\ref{fig:main-summary} aggregates the main results across the five target generators and three datasets. \methodname{} achieves the highest mean ASR across all evaluated conflict-resolution methods, showing that its advantage is consistent rather than driven by a particular generator or dataset.

\paragraph{Retrieval-Conditioned Attack Effectiveness.}
Table~\ref{tab:conditional_asr} reports $\mathrm{ASR}_{\mathrm{strict}}$/ASR conditioned on the poisoned document appearing in the top-5, averaged across the five target generators. \methodname{} achieves the highest conditional ASR in all nine conflict-resolution settings and the highest conditional $\mathrm{ASR}_{\mathrm{strict}}$ in eight of them. This confirms that its advantage persists after successful retrieval and is not primarily attributable to retrieval success.
\begin{table}[h]
\centering
\small
\newcommand{\condmetric}[1]{{\fontsize{9pt}{8.5pt}\selectfont #1}}
\setlength{\tabcolsep}{3.2pt}
\renewcommand{\arraystretch}{1.0}
\begin{tabular}{lcccc}
\toprule
Resolution & PRAG & Auth. & PARA. & Ours \\
\midrule

\rowcolor{gray!15}
\multicolumn{5}{c}{\textbf{NQ}} \\
Vanilla
& \condmetric{50.9/56.2}
& \condmetric{51.5/57.3}
& \condmetric{\textbf{86.2/87.8}}
& \condmetric{85.8/87.7} \\
Astute
& \condmetric{7.3/7.3}
& \condmetric{15.9/17.0}
& \condmetric{21.7/23.0}
& \condmetric{\textbf{41.9/45.1}} \\
Faithful
& \condmetric{50.1/51.7}
& \condmetric{60.6/61.2}
& \condmetric{88.2/88.5}
& \condmetric{\textbf{94.5/94.7}} \\
MADAM
& \condmetric{11.3/17.6}
& \condmetric{12.2/22.5}
& \condmetric{18.2/34.8}
& \condmetric{\textbf{26.7/46.7}} \\

\midrule
\rowcolor{gray!15}
\multicolumn{5}{c}{\textbf{HotpotQA}} \\
Vanilla
& \condmetric{54.8/57.4}
& \condmetric{66.1/70.1}
& \condmetric{\textbf{88.2/88.4}}
& \condmetric{86.0/88.0} \\
Astute
& \condmetric{13.4/13.6}
& \condmetric{17.3/18.3}
& \condmetric{36.6/36.8}
& \condmetric{\textbf{44.8/45.6}} \\
Faithful
& \condmetric{55.6/56.4}
& \condmetric{64.1/64.7}
& \condmetric{87.2/87.2}
& \condmetric{\textbf{87.4/88.0}} \\
MADAM
& \condmetric{12.8/17.6}
& \condmetric{13.3/19.5}
& \condmetric{\textbf{20.2}/30.0}
& \condmetric{19.4/\textbf{32.4}} \\

\midrule
\rowcolor{gray!15}
\multicolumn{5}{c}{\textbf{MS-MARCO}} \\
Vanilla
& \condmetric{35.5/36.9}
& \condmetric{49.7/53.0}
& \condmetric{72.9/74.0}
& \condmetric{\textbf{82.7/84.9}} \\
Astute
& \condmetric{4.9/5.1}
& \condmetric{9.9/10.5}
& \condmetric{18.8/19.4}
& \condmetric{\textbf{36.2/38.1}} \\
Faithful
& \condmetric{39.6/40.2}
& \condmetric{66.8/67.1}
& \condmetric{78.0/78.0}
& \condmetric{\textbf{90.7/91.5}} \\
MADAM
& \condmetric{6.3/15.1}
& \condmetric{8.3/18.1}
& \condmetric{11.1/21.6}
& \condmetric{\textbf{12.9/31.5}} \\
\bottomrule
\end{tabular}

\caption{Attack effectiveness conditioned on successful retrieval. Each cell reports conditional $\mathrm{ASR}_{\mathrm{strict}}$/ASR (\%), conditioned on $d_{\mathrm{adv}}\in\mathrm{Top}\text{-}5$ and averaged across the five target generators.}
\label{tab:conditional_asr}
\end{table}

\paragraph{Cross-Attacker Generalization.}
Table~\ref{tab:attack_model_results} reports the complete NQ results when the attack pipeline is instantiated with five different attack models.
\begin{table}[h]
\centering
\small
\newcommand{\metric}[1]{{\fontsize{8.0pt}{8.5pt}\selectfont #1}}
\setlength{\tabcolsep}{2.2pt}
\renewcommand{\arraystretch}{1.0}
\begin{tabular}{llcccc}
\toprule
\multirow{2}{*}{Atk Model} & \multirow{2}{*}{Resolution}
& \multicolumn{4}{c}{Attack Methods} \\
\cmidrule(lr){3-6}
& & PRAG & Auth. & PARA. & \textit{Ours} \\
\midrule

\multirow{4}{*}{DeepSeek}
& Vanilla
& \metric{45/51/52} & \metric{49/42/42} & \metric{28/69/70} & \metric{\textbf{16}/\textbf{82}/\textbf{82}} \\
& Astute
& \metric{83/11/11} & \metric{70/22/22} & \metric{74/24/26} & \metric{\textbf{40}/\textbf{58}/\textbf{61}} \\
& Faithful
& \metric{40/54/55} & \metric{32/52/52} & \metric{16/73/74} & \metric{\textbf{14}/\textbf{80}/\textbf{80}} \\
& MADAM
& \metric{\textbf{55}/19/23} & \metric{61/20/32} & \metric{57/25/49} & \metric{\textbf{55}/\textbf{39}/\textbf{64}} \\
\midrule

\multirow{4}{*}{GPT}
& Vanilla
& \metric{60/39/40} & \metric{23/74/74} & \metric{41/55/56} & \metric{\textbf{15}/\textbf{83}/\textbf{83}} \\
& Astute
& \metric{85/11/13} & \metric{81/15/15} & \metric{84/12/13} & \metric{\textbf{41}/\textbf{55}/\textbf{56}} \\
& Faithful
& \metric{50/42/42} & \metric{18/79/79} & \metric{31/61/61} & \metric{\textbf{13}/\textbf{80}/\textbf{80}} \\
& MADAM
& \metric{58/11/15} & \metric{60/28/45} & \metric{64/20/35} & \metric{\textbf{53}/\textbf{36}/\textbf{57}} \\
\midrule

\multirow{4}{*}{Gemini}
& Vanilla
& \metric{34/65/68} & \metric{24/74/74} & \metric{\textbf{18}/\textbf{80}/\textbf{80}} & \metric{22/73/73} \\
& Astute
& \metric{83/12/13} & \metric{85/9/9} & \metric{73/23/24} & \metric{\textbf{57}/\textbf{35}/\textbf{35}} \\
& Faithful
& \metric{36/63/64} & \metric{20/76/77} & \metric{\textbf{9}/\textbf{85}/\textbf{85}} & \metric{22/72/72} \\
& MADAM
& \metric{\textbf{54}/20/27} & \metric{58/27/42} & \metric{56/29/47} & \metric{57/\textbf{33}/\textbf{61}} \\
\midrule

\multirow{4}{*}{Llama}
& Vanilla
& \metric{67/24/26} & \metric{48/49/52} & \metric{18/\textbf{79}/\textbf{79}} & \metric{\textbf{17}/69/69} \\
& Astute
& \metric{85/7/7} & \metric{88/6/7} & \metric{79/16/17} & \metric{\textbf{66}/\textbf{25}/\textbf{25}} \\
& Faithful
& \metric{55/28/29} & \metric{44/54/56} & \metric{12/\textbf{85}/\textbf{85}} & \metric{\textbf{11}/68/68} \\
& MADAM
& \metric{60/3/5} & \metric{64/26/37} & \metric{\textbf{53}/\textbf{27}/\textbf{45}} & \metric{\textbf{53}/23/44} \\
\midrule

\multirow{4}{*}{Qwen}
& Vanilla
& \metric{34/64/68} & \metric{24/74/74} & \metric{\textbf{17}/74/74} & \metric{20/\textbf{76}/\textbf{76}} \\
& Astute
& \metric{80/17/25} & \metric{84/11/11} & \metric{72/22/25} & \metric{\textbf{55}/\textbf{42}/\textbf{43}} \\
& Faithful
& \metric{36/63/65} & \metric{19/\textbf{75}/\textbf{75}} & \metric{\textbf{15}/68/69} & \metric{18/73/73} \\
& MADAM
& \metric{\textbf{55}/21/34} & \metric{60/28/46} & \metric{59/26/\textbf{51}} & \metric{57/\textbf{30}/\textbf{51}} \\
\bottomrule
\end{tabular}
\caption{Results across attack models on NQ. Attack columns report
ACC$\downarrow$/$\mathrm{ASR}_{\mathrm{strict}}$$\uparrow$/ASR$\uparrow$.}
\label{tab:attack_model_results}
\end{table}

Across the 20 probing-LLM and RAG-setting combinations, \methodname{} achieves the highest ASR in 14 cells. Of the remaining six, PARADOX leads in five and AuthChain in one. All six occur under vanilla RAG or FaithfulRAG, the same settings in which these baselines are already most competitive in the main experiments. The exceptions are therefore concentrated in particular RAG settings rather than associated with a systematic failure under any probing model.

Performance nevertheless varies with the probing LLM. DeepSeek-V3.2 and GPT-5.2 achieve the highest mean ASR at 71.3 and 69.0, followed by Qwen3.5-Plus and Gemini-3-Flash at 60.8 and 60.3, while Llama-3.3-70B reaches 51.5. This pattern is consistent with \methodname{}'s reliance on knowledge elicitation: a more informative elicited belief $\mathcal{F}_q$ can provide stronger compatibility constraints for constructing the pivot event. Thus, the probing model affects the strength of the resulting attack, but \methodname{}'s advantage is not tied to a single model.

\subsection{Full Analysis}
\subsubsection{Main result.}
\paragraph{Overall: \methodname{} attains the strongest attack across cells and generators.} \methodname{} achieves the highest ASR in 35 of 45 conflict-resolution cells in Table~\ref{tab:main_result} and yields the largest mean ASR on every resolution method, exceeding the strongest prior baseline by $+14.7$, $+6.5$, and $+7.9$ on AstuteRAG, FaithfulRAG, and MADAM-RAG, respectively (Figure~\ref{fig:main-summary}). \methodname{} also produces the largest mean ACC drop overall, leading the strongest prior baseline by $+6.6$ points ($29.3$ vs.\ $22.7$). The lead holds across all five generators, including GPT-5.2, the most attack-resistant of the five and precisely the kind of broad-parametric-coverage model that \methodname{} is designed to exploit.
 
\begin{itemize}
\item\textbf{AstuteRAG: parametric arbitration is bypassed.} Prior attacks rarely raise ASR above 20\% on AstuteRAG, whose explicit parametric arbitration rejects documents that contradict the model's knowledge. \methodname{} reaches a mean ASR of $38.4$ and nearly doubles the mean ACC drop of the strongest prior attack ($28.3$ vs.\ $15.2$), with the Llama-3.3-70B + NQ cell showing a 73-point collapse ($90 \to 17$).
 
\item\textbf{FaithfulRAG: \methodname{} and PARADOX are comparable, with \methodname{} stronger on average.} On FaithfulRAG, \methodname{} attains the higher ASR on 11 of 15 cells and the higher mean ($78.7$ vs.\ $72.2$ for PARADOX). PARADOX is comparable here but trails \methodname{} by $14.7$ and $7.9$ mean ASR points on AstuteRAG and MADAM-RAG.
 
\item
\textbf{MADAM-RAG: gains shift from suppression to hedging.} Multi-agent debate is orthogonal to parametric knowledge, and accuracy moves little under any attack (mean ACC drop $\leq 6.0$). \methodname{} nonetheless attains the highest mean ASR ($32.8$ vs.\ $24.9$) and routes more outputs into hedging, with $6/15$ cells satisfying $\mathrm{ACC} + \mathrm{ASR} > 100$.
\end{itemize}

\paragraph{The advantage narrows when retrieved content is consumed without scrutiny.} On vanilla RAG, \methodname{} attains the highest ASR on $10$ of $15$ cells in Table~\ref{tab:main_result}, with prior attacks closing the remaining five (e.g., GPT-5.2 + HotpotQA: $17/84$ for \methodname{} against $4/96$ for PARADOX in that cell). The mean-ASR margin of \methodname{} over the strongest competing attack shrinks to $+4.9$, against $+14.7$, $+6.5$, and $+7.9$ on AstuteRAG, FaithfulRAG, and MADAM-RAG respectively. This pattern isolates the source of our gain. \methodname{}'s pivot event is engineered to be persuasive: it preserves what the system already accepts as true, and routes the target answer through a coherent, plausibly-sourced novel development. Such persuasiveness pays off only when the system actually weighs the credibility of what it reads, whether by arbitrating against parametric knowledge, validating individual facts, or debating across passages. Vanilla RAG performs none of these and consumes any retrieved document at face value; in this regime, a brute-force attack that simply asserts $\tilde{y}$ in retrievable text is already sufficient, and \methodname{}'s additional design effort yields little marginal return. Even so, \methodname{} retains the strongest attack on most cells, consistent with its dominance across the three conflict-resolution settings.

\paragraph{Linguistic Fluency.}
We assess linguistic fluency as a surface-level aspect of attack stealthiness using per-sample perplexity (PPL), a standard fluency proxy that underlies common low-quality-text filters~\citep{zou-Usenix2025-poisonedrag}.
Across all three scorers in Panel B of Table~\ref{tab:main_result}, \methodname{} sits in the same fluency tier as AuthChain, the strongest stealthy prior baseline, which is roughly $2\times$ more natural than PARADOX and PRAG. \methodname{} achieves this parity through a single feed-forward pipeline, without AuthChain's iterative judge and revise refinement loop. The wider \methodname{}--AuthChain gap on GPT-2 ($+37\%$, vs.\ $+4\%$ on Mistral) reflects that scorer's pre-2019 training data and limited exposure to recent-announcement language, where our reproductions of AuthChain and PoisonedRAG also fall within $1.5\%$ and $6\%$ of the originally reported GPT-2 PPL~\citep{chang-EMNLP2025findings-authchain}.

\section{Analysis Through Case Study}
\label{sec:analysis}

The same query reveals a different failure path through each resolution method (Box~\ref{box:test213}); tracing the path shows that each system fails for an architectural reason rather than from being deceived.

\begin{tcolorbox}[enhanced, float=t, width=\linewidth, colback=gray!3, colframe=gray!55, boxrule=0.5pt, arc=3pt, leftrule=2pt, drop shadow={black!22, xshift=1pt, yshift=-1pt, opacity=0.45}, title=\textbf{(NQ test split, sample 213) \\Question: ``Who won the most NBA MVP awards?''}, fonttitle=\small, coltitle=black, colbacktitle=gray!18, fontupper=\small, label={box:test213}]
\textbf{Gold:} Kareem Abdul-Jabbar\\  \textbf{Target ($\tilde{y}$):} Michael Jordan \\ \textbf{Pivot event:} a fabricated \emph{2024 NBA Official History Database} revision and AP re-vote project that reassigns Kareem's 1977 MVP to Jordan.

\medskip
\hrule height 0.3pt
\medskip

\noindent\textbf{AstuteRAG.}\; \emph{Parametric dump:} ``Kareem Abdul-Jabbar\,\dots\,\textbf{six (6) regular season MVP awards}. Confidence Level: High. Well-documented and undisputed.'' $\;\rightarrow\;$ \textbf{Output:} \emph{Michael Jordan}.

\smallskip
\noindent\textbf{FaithfulRAG.}\; \emph{Self-facts:} ``Kareem won six MVP'', ``Jordan has five''. \emph{[Final Verification]:} ``\emph{Context provides updated information from a \textbf{2024 revision}, confirming Jordan now holds the record\,\dots\,after corrections}; no conflict.'' $\;\rightarrow\;$ \textbf{Output:} \emph{Michael Jordan}.

\smallskip
\noindent\textbf{MADAM-RAG.}\; \emph{R1 aggregator:} ``Agent~1 [PURPOSE] claims Jordan based on a \emph{hypothetical revision not supported by widely accepted NBA history}.'' \emph{R2--3:} four agents converge on Kareem; Agent~1 alone holds Jordan; aggregator switches to \textsc{Both}. $\;\rightarrow\;$ \textbf{Output:} \textsc{Both} Jordan and Kareem.
\end{tcolorbox}

\noindent\textbf{AstuteRAG} extracts parametric belief and enters consolidation with the gold answer already in hand---on test213 the dump explicitly states that Kareem holds six MVPs with high confidence. Consolidation, however, is a generative step rather than a logical arbiter: it reads the retrieved passage not as a candidate to be vetted against the dump but as additional evidence to be folded into the final answer. \methodname{}'s pivot event is admissible into that folding because it never contradicts the dump---it presents itself as a procedurally legitimate update that supersedes it. The arbitration succeeds at its prerequisite (eliciting the truth) and fails at the step it was designed to perform (using that truth to reject what conflicts with it).

\noindent\textbf{FaithfulRAG} performs an explicit conflict-detection step that checks each retrieved passage for inconsistency with the model's self-facts. The check is semantic, not logical---it asks whether the passage contradicts a fact, not whether it competes with one. On test213 the self-facts correctly hold both ``Kareem won six'' and ``Jordan has five'', and \methodname{}'s pivot event neither denies nor revises either statement; it adds a third (a 2024 reassignment) that is logically consistent with both. The detector therefore returns its honest verdict, no conflict, and the resolver designed to catch parametric--retrieval contradictions cannot recognize that a non-contradicting addendum may still be a fabrication.

\noindent\textbf{MADAM-RAG} resolves disagreement through inter-agent consensus rather than through the merits of each agent's evidence. On test213 the round-1 aggregator initially flags Agent~1's revision as ``a hypothetical revision not supported by widely accepted NBA history''---an explicit identification of the poison. Once later rounds reveal that Agent~1 will not retract while four other agents converge on Kareem, the aggregator has no mechanism for breaking the impasse on grounds of evidential weight; it defaults to representing the persistent disagreement as \textsc{Both} answers being correct. \methodname{} therefore does not need to win the debate---only to anchor a single agent's claim long enough for the aggregation to interpret unresolved minority dissent as legitimate plurality.



\end{document}